\documentstyle[12pt]{article}

\begin{document}

\baselineskip 16pt

\title{
DEFENDING  TIME-SYMMETRIZED QUANTUM THEORY}
\author{ Lev Vaidman}
\date{}
\maketitle

\begin{center}
{\small \em School of Physics and Astronomy \\
Raymond and Beverly Sackler Faculty of Exact Sciences \\
Tel Aviv University, Tel-Aviv 69978, Israel. \\}
\end{center}

\vspace{2cm}
\begin{abstract}
  Recently, several authors have criticized time-symmetrized quantum
  theory originated by the work of Aharonov et al. (1964).  The core
  of this criticism was the proof, which appeared in various forms,
  showing that counterfactual interpretation of time-symmetrized
  quantum theory cannot be reconciled with the standard quantum
  theory.  I argue here that the apparent contradiction appears due to
  inappropriate usage of traditional time asymmetric approach to
  counterfactuals, and that the contradiction disappears when the
  problem is analyzed in terms of time-symmetric counterfactuals. I
  analyze various aspects of time-symmetry of quantum theory and
  defend the time-symmetrized formalism.
\end{abstract}

\vfill\break

 \noindent {\bf 1. Introduction.~~ } I shall discuss a pre- and
post-selected quantum system, i.e.  measurements performed at the time
between two other measurements.  The time-symmetric formalism for the
description of such systems was proposed by Aharonov, Bergmann, and
Lebowitz (ABL) (1964) and was developed in recent years. A partial
list of references includes Aharonov et al.  (1985), Aharonov and
Vaidman (1990, 1991).  Several authors criticized the time-symmetric
approach to quantum theory in general and some of its particular
applications. The most representative example is the work of Sharp and
Shanks (1993). They presented a proof, which was later repeated and
used by others, that the counterfactual interpretation of the ABL
probability rule (Eq. 2 below), cannot be reconciled with the standard
quantum theory. I shall claim here that the proof contains an error
since it presuppose time asymmetry in order to reach a contradiction
with time symmetry. The asymmetry was implicitly assumed through the
conventional approach to counterfactual statements. I argue that for
analyzing experiments on pre- and post-selected quantum system one
should use time-symmetric counterfactuals and then no contradiction
arises.

The plan of this work is as follows.  In Section 2 I shall present a
brief review of the time-symmetrized formalism and in Section 3 a
brief review of the concept of counterfactual. In Section 4 I analyze
possible counterfactual interpretations of the ABL rule.  Section 5 is
devoted to the analysis of the inconsistency proof of Sharp and Shanks
and its variations.  In section 6 I discuss related time asymmetry
preconceptions in quantum theory. In section 7 the time symmetry (and
asymmetry) of the process of quantum measurement is analyzed in order
to give a rigorous context to the previous discussion about the time
symmetry of the ABL rule. An application of the time-symmetrized
approach which allows the definition of new concepts (which I call
``elements of reality'') is considered in section 8. Section 9
concludes the paper with a brief summary and some discussion of the
time-symmetrized quantum theory in the framework of the many-worlds
interpretation.

\vskip 1.cm \noindent
{\bf  2. Time-Symmetrized Formalism.~~}
In standard quantum theory a complete description of a system at a
given time is given by a quantum state $|\Psi \rangle$. It yields the
probabilities for all outcomes $a_i$ of a measurement of any variable
$A$ according to the equation
\begin{equation}
  \label{prob1}
  {\rm Prob}(a_i) = |\langle \Psi | {\bf P}_{A=a_i} | \Psi \rangle |^2
\end{equation}
where ${\bf P}_{A=a_i}$ is the projection operator on the subspace defined
by $A= a_i$.  Eq. 1 is intrinsically asymmetric in time: the state
$|\Psi\rangle$ is determined by some measurements in the past and it
evolves toward the future.  The time evolution between the
measurements, however, is considered time symmetric since it is
governed by the Schr\"odinger equation for which each forward evolving
solution has its counterpart (its complex conjugate with some other
well understood simple changes) evolving backward in time. The
asymmetry in time of the standard quantum formalism is manifested in the
absence of  the quantum state evolving backward in time from 
future measurements (relative to the time in question).

Time-symmetrized quantum theory describes a system at a given time by
a two-state vector $\langle \Psi_2| |\Psi_1 \rangle$. It yields the
(conditional) probabilities for all outcomes $a_i$ of a measurement of
any variable $A$ according to the generalization of the ABL formula
(Aharonov and Vaidman, 1991): \begin{equation}
  \label{ABL}
 {\rm Prob}(a_i) = {{|\langle \Psi_2 | {\bf P}_{A=a_i} | \Psi_1
\rangle |^2} \over{\sum_j|\langle \Psi_2 | {\bf P}_{A=a_j} | \Psi_1
\rangle |^2}} .  \end{equation} The time symmetry does not mean that a
system described by the two-state vector $\langle \Psi_2| |\Psi_1
\rangle$ is identical, in regard to its physical properties, to a
system described by the two-state vector $\langle \Psi_1| |\Psi_2
\rangle$.\footnote{Note,
  however, that if we limit ourselves to ``physical properties'' which
  are results of standard
  ideal measurements (whose probabilities are governed by Eq. 2), this
  symmetry property holds too.} The time symmetry means that $\langle
\Psi_2|$ and $|\Psi_1 \rangle$ enter the equations, and thus govern
the observable results, on equal footings. For example, (almost)
standard measurement procedure with weakened coupling (which we call
{\em weak
  measurement}, Aharonov and Vaidman 1990) yields {\em weak values}
defined as \begin{equation} A_w \equiv { \langle{\Psi_2} \vert A
\vert\Psi_1\rangle \over \langle{\Psi_2}\vert{\Psi_1}\rangle } .
\label{wv} \end{equation} When we interchange $\langle \Psi_2|$ and
$|\Psi_1 \rangle$ the weak value changes to its complex conjugate and
this can be observed in certain experiments.  In Eq. 2 and in Eq. 3
the two states enter on the same footing and the legitimacy of the
application of these equations (especially of Eq. 2) is what I defend
in this paper.

In order to explain how to obtain a quantum system described at a
given time $t$ by a two-state vector $\langle \Psi_2| |\Psi_1 \rangle$
we shall assume for simplicity that the free Hamiltonian of the system
is zero. In this case it is enough to prepare the system at time $t_1$
prior to time $t$ in the state $|\Psi_1\rangle$, to ensure no
disturbance between $t_1$ and $t$ as well as between $t$ and $t_2$,
and to find the system at $t_2$ in the state $|\Psi_2\rangle$. It is
crucial that $t_1 < t < t_2$, but the relation between these times and
``now'' is not fixed. The times $t_1, t, t_2$ might all be in the
past, or we can discuss future measurements and then they are all in
the future; we just have to agree to discard all cases when the
measurements at time $t_2$ does not yield the outcome corresponding to
the state $|\Psi_2\rangle$.

Note the asymmetry between the measurement at $t_1$ and the
measurement at $t_2$. Given an ensemble of quantum systems, it is
always possible to prepare all of them in a particular state
$|\Psi_1\rangle$, but we cannot ensure finding the system in a
particular state $|\Psi_2\rangle$.  Indeed, if the pre-selection
measurement yielded a result different from projection on
$|\Psi_1\rangle$ we can always change the state to $|\Psi_1\rangle$,
but if the measurement at $t_2$ did not show $|\Psi_2\rangle$, our
only choice is to discard such a system from the ensemble. Note also
the asymmetry of the measurement procedures. The measurement device
has to be prepared before the measurement interaction in the ``ready''
state and we cannot ensure finding the ``ready'' state after the
interaction. We might use some intermediate system which interacts
with the observed system such that this intermediate system has the
essential symmetry of the states before and after the interaction. But
then the problem will move to the next level of the measurement
procedure chain, and it always reaches the asymmetry, because,
according to the definition of measurement, the observer does not know
the result before the interaction but he does, after the
measurement. These asymmetries, however, are not relevant to the
problem we consider here.  We study the symmetry relative to the
measurements at time $t$ for a given pre- and post-selected system,
and we do not investigate the time-symmetry of obtaining such a
system. The only important detail is that the measurement coupling at
time $t$ has to be time-symmetric, as is assumed in ideal quantum
measurements. See more discussion below, in section 7.

\vskip 1.cm \noindent {\bf 3.Counterfactuals.~~} There are many
philosophical discussions on the concept of counterfactuals and
especially on the time's arrow in counterfactuals.  Probably, the
theory of counterfactuals of Lewis (1973) receives the most
attention. Based on his theory Lewis (1986) discusses the asymmetry of
counterfactuals between past and future. He analyses mainly
deterministic worlds and claims that indeterminism, in particular the
indeterminism of the process of reduction of a quantum state in the
process of quantum measurement, does not lead to an asymmetry:
\begin{quotation} If there is a process of reduction of the wave
packet in which a given superposition may be followed by any of many
eigenstates, equally this is a process in which a given eigenstate may
have been preceded by any of many superpositions. Again we have no
asymmetry. (1986, 39) \end{quotation} I disagree with this
argument. The superposition which precedes the measurement is uniquely
defined by the classical records regarding measurements in the
past. In any way, the number of possible superpositions is not
comparable with the number of eigenstates, so no symmetry can be seen
here.

 Apart from the possible connection between indeterminism and an
apparent time asymmetry of our world, the importance of the
indeterminism of the standard quantum theory is that it opens room for
counterfactual questions about results of measurements without
involving ``miracles'', i.e. events in which physical laws breaks
down.  Although Lewis devotes a large part of his theory to
considering these ``miracles'' which are irrelevant for our
discussion, I do adopt the basic approach of his analyses of
counterfactuals, i.e. the usage of the language of ``possible
worlds''. It is not clear if this is the only way to go, but I find it
fruitful and certainly legitimate.

Another important work on the subject was done by Bennett (1984).
He reaches the conclusion (with which I tend to agree) that Lewis
failed to {\em derive} the temporal asymmetry of counterfactuals from
general principles. Bennett develops his ``Unified Symmetric Theory''
of counterfactuals. It is based on the concept of ``$T$-closest
$P$-world'' which is the world ``closest'' (whatever it means) to the
actual world at time $T$ at which the proposition $P$ is true.

There are many discussions of   counterfactuals in quantum theory, mostly
in the  context of EPR-Bell type experiments. Some of the examples are
Skyrms (1982),  Peres (1993),  Mermin (1989) (which, however, does not
use the word counterfactual), and Bedford and  Stapp (1995) who even
present an analysis of a Bell-type argument in the formal language of
the Lewis theory of counterfactuals. The common situation is that a
composite system is described at a certain time by some entangled state
and then an array of incompatible measurements on this system at a
later time is considered. Various conclusions are derived from
statements about  the results of these measurements. Since these
measurements are incompatible they cannot be all performed together,
so it must be that at least some of them were not actually performed.
This is why they are called counterfactual statements. Note that there is
no requirement that none of them are performed, although it might be
so. The actual world is specified by its state at the initial time so
it fits the general framework of Bennett's theory of counterfactuals.

In the situations discussed in this work the actual world is specified
by its state at two times. Thus, Bennett's basic concept of $T$-closest
$P$-world cannot be applied directly, but the following quotation of
Bennett seems very relevant:
\begin{quotation}
  Here is an easier example. At $T_1$ I bet that when the coin is
  tossed at $T_2$ it will come up heads; and in the upshot it does
  just that; but this is a purely chance event, with no causally
  sufficient prior conditions. Now consider the conditional ``If I had
  bet on tails at $T_1$ I would have lost.'' Everyone I have polled is
  inclined to say that that conditional is true, despite the fact that
  at some of the $T_1$-closest ``I bet on tails'' worlds the coin
  comes up heads [tails?\footnote{It seems to me that there is some
  mismatch between ``tails'' and ``heads'' which I, hopefully,
  corrected in the brackets.}] at $T_2$. (Why does it come up heads
  [tails] at some of those worlds? Because, since the fall of the coin
  had no causally sufficient prior conditions, every ``tails''
  [``heads''] world is indistinguishable, in respect of its state at
  $T_1$, from some ``heads'' [``tails''] world.)  If I am to respect
  these judgments I must modify my theory,... (1984, 76)
\end{quotation}
In his modification of the theory (which he only sketches in a few
lines) there is no formal symmetry between the times $T_1$ and $T_2$.
But this is not because Bennett introduces temporal asymmetry here,
but because  the  two times have different status.
 The time $T_1$ has
a special status as the one at which various possibilities (the type
of bet to be taken) have to be chosen, and this is why Bennett relates
the concept of $T$-closest worlds to $T_1$ and not to $T_2$. In order to
analyze temporal symmetry we have to consider a symmetric setup. Let
us add another   coin tossing  at time $T_0$ prior to
the time of the bet. Unquestionably, ``everyone Bennett has polled''
who suggested we accept that in all relevant (counterfactual) worlds
the outcome of the coin tossing  at time $T_2$ must coincide with that
in the actual world would also suggest that the outcome of the
coin tossing  at time $T_0$ must coincide with that in the actual world.
Thus, we can see that even a modified theory of counterfactuals of
Bennett, which is suitable for the analysis of possible measurements
performed between two other measurements, is symmetrical with regard to
the past and the future of time $T_1$.

The example of Bennett is not identical to the problem of measurement
performed on quantum system at the time between two other
measurements. In Bennett's case there was no physical mechanism
according to which the decision of which bet to make could influence
the result of the coin toss.\footnote{Moreover, one can make an
experiment which will show that the bet decision does not change the
probability of the result of the coin toss.} In contrast, most
intermediate quantum measurements change the probabilities for the
results of the later measurement.  However, it seems to me that the
theory of counterfactuals which respects the actual event at time
$T_2$ in spite of the fact that it should be identical with a specific
(actual) result only in 50\% cases is much more close to the theory of
counterfactuals accepting an actual event at time $T_2$ when the
probability is different and even influenced by the intermediate
action, rather than to the theory which disregards the actual result
at time $T_2$, as is suggested in alternative interpretations.

\vskip 1.cm \noindent {\bf 4. Counterfactual Interpretations of the
  ABL Probability Rule.~~} In this section I shall consider three ways
to interpret the ``counterfactual interpretation''. The first
interpretation I cannot comprehend, but I have to discuss it since it
was proposed and used in the criticism on the time-symmetrized quantum
theory. I believe that I understand the meaning of the second
interpretation, but I shall argue that it is  not appropriate
for the problem which is discussed here. The last interpretation is
the one I want to adopt and I shall present several arguments in its
favor.

\vskip .2cm
\noindent
{\bf Interpretation (a)~~}
{\it  Counterfactual probability as the probability of the result of a
measurement which has not been performed.}

Let me quote Sharp and Shanks: 
\begin{quotation}
  ...for, conditionalizing upon specified results of measurements of
  $M_I$ and $M_F$, there is no reason to assign the same values to the
  following probabilities: the probability that an intervening
  measurement of $M$ had the result $m^j$ given that such a
  measurement in fact took place, and the probability that intervening
  measurement {\em would have had} the result $m^j$ given that no such
  intervening measurement of $M$ in fact took place. In other worlds
  there is no reason to identify ${\rm Prob}(M= m^j|{\rm \bf
    E}_M[\psi^i_I, \psi^k_F])$ and ${\rm Prob}(M= m^j|{\rm \bf
    E}[\psi^i_I, \psi^k_F])$.(1993, 491)
\end{quotation}
I can not comprehend the  meaning of the probability for the result
$M=m^j$ given that the measurement $M$ has not take place. As far as I
can see $ {\rm Prob}(M= m^j|{\rm \bf  E}[\psi^i_I, \psi^k_F])$ has no
physical meaning. Sharp and Shanks  continue:
\begin{quotation}
 (For a classical illustration, consider a drug which, if injected to
 facilitate a medical test at $t$, has an effect, starting shortly after
 the test and persisting past $t_F$, on the the value of the tested
 variable. Suppose that it is unknown whether a test was conducted at
 $t$, but that a value for the tested variable is obtained at $t_F$. Using
 the value  at $t_F$, we would estimate differently the value prior to
 $t$ depending on whether we assume that a test did or did not take
 place at $t$.)
\end{quotation}
This might explain what they have in mind, but the argument does not
hold since in many situations there is no  quantum
mechanical counterpart to the classical case of ``the value [of a tested
variable] prior to  $t$'' . In standard quantum
theory {\em unperformed experiments have no results}, see Peres (1978).

Cohen and Hiley  partially acknowledge the problem admitting
that at least in the framework of the orthodox interpretation this is
meaningless concept:
\begin{quotation}
  In other words we cannot necessarily assume that the ABL rule will
  yield the correct probabilities for what the results of the
  intermediate measurements {\em would} have been, {\em if} they had
  been carried out, in cases where these measurements have not {\em
    actually} been carried out. In fact, this sort of counterfactual
  retrodiction has no meaning in the orthodox (i.e., Bohrian)
  interpretation of quantum mechanics, although it can legitimately be
  discussed within the standard interpretation [von Neumann (1955)] and
  within some other interpretations of quantum mechanics (see, for
  example, Bohm and Hiley[1993]).(1996, 3)
\end{quotation}
I fail to understand the interpretation (a) in any framework.
Maybe, if we restrict ourselves to the cases in which the system at
the intermediate time is in an eigenstate of the variable which we
intended to measure, (but we had not), we can associate the
probability 1 with such unperformed measurements. This is close to
the idea of Cohen (1995) to consider counterfactuals in the restricted
cases corresponding to consistent histories introduced by Griffith
(1984). But, as far as I can see, interesting situations do not
correspond to consistent histories, and therefore no novel (relative
to classical theory) features of quantum theory can be seen in this
way.  It is possible that what Cohen and Hiley (1996) have in
mind is the interpretation (b) which I shall discuss next.

\vskip .2cm
\noindent
{\bf Interpretation (b)}

{\it Counterfactual probability as the probability of the result of a
measurement would it have been performed based on the information
about the world in which the measurement has not been performed.}

At time $t_1$ we preselect the state $|\Psi_1\rangle$. We do not
perform any measurement at time $t$. We perform a measurement at time
$t_2$ and find the state $|\Psi_2\rangle$. We ask, what would be the
probability for the results of a measurement performed at time $t$ in
a world which is identical to the actual world at time $t_1$.

This is a meaningful concept, but I believe that it is not adequate
for discussing pre- and post-selected quantum systems because it is
explicitly asymmetric in time. The counterfactual world is identical to the
actual world at time $t_1$ and might not be identical at time $t_2$.

This interpretation of the ABL rule is  clearly inconsistent with
predictions of quantum theory. According to the orthodox or standard
interpretation the information about the actual world at  time $t_2$ is
irrelevant since the state $|\Psi_1\rangle$ together with the
requirement of no disturbance between $t_1$ and $t$ define completely
the probabilities of all possible measurements at time $t$. Therefore,
the ABL formula for probabilities which includes explicitly dependence
on the result of the measurement at time $t_2$ cannot be consistent
with quantum theory.

One may speculate about possible modifications of quantum theory which
reconstruct statistical predictions of standard quantum theory but
include, in addition, some hidden variables which specify individual
outcomes of seemingly random results of quantum measurements. Then the
information about the results of the measurements at time $t_2$ in a
run of the experiment
without an intermediate measurement might add to the description of the
actual world at  time $t_1$. See, for example, the discussion about
such a situation in the framework of the Bohm (1952) theory by Aharonov
and Albert (1987).  In the framework of a hidden variable theory, for
estimating the probabilities of the result of measurements at time
$t$ we have to consider the sub-ensemble of the pre-selected systems at
time $t_1$ which have hidden variables corresponding to the appropriate
result of the measurement at time $t_2$ on the condition that no
disturbance (and, in particular no measurement at time $t$) took place
between $t_1$ and $t_2$.  The question of consistency between the ABL
rule and the predictions of quantum theory in such framework seems to
be a nontrivial problem.  One may just recall the difficulty in the
framework of the Bohm (1952) hidden variable theory according to which
the outcome of a spin measurement might depend not only on the hidden
variable of the system, but also on the state of the measuring device,
(see Albert 1992, 153-154). However, I shall present now a simple
example which allows us to show the inconsistency between quantum theory
and this interpretation of the ABL rule irrespectively of the details
of the hidden-variable theory.

Consider a spin-1/2 particle pre-selected at time $t_1$ in the state
$|\uparrow_z\rangle$ and post-selected at the time $t_2$ in the same
state $|\uparrow_z\rangle$. We ask what is the probability for finding
spin ``up'' in the direction $\hat \xi$ which makes an angle $\theta$
with the direction $\hat z$, at the intermediate time $t$. In this
case, the hidden variables, even if they exist, cannot change that
probability because any particle pre-selected in the state
$|\uparrow_z\rangle$, irrespectively of its hidden variable, yields
the outcome ``up'' in the post-selection measurement at time
$t_2$. Therefore, the statistical predictions about the intermediate
measurement at time $t$ must be the same as for the pre-selected only
ensemble (these are {\em identical} ensembles in this case), i.e.
\begin{equation}
  \label{abl-qm}
  {\rm Prob}(\uparrow_\xi) =|\langle \uparrow_\xi |
\uparrow_z\rangle|^2 = \cos^2(\theta/2).
\end{equation}
The ABL formula, however, yields: 
\begin{equation}
  \label{abl-xiz}
  {\rm Prob}(\uparrow_\xi) = {{|\langle \uparrow_z |
      {\bf P}_{\uparrow_\xi} | \uparrow_z \rangle |^2}\over{|\langle
      \uparrow_z | {\bf P}_{\uparrow_\xi} | \uparrow_z \rangle |^2
    +|\langle \uparrow_z | {\bf P}_{\downarrow_\xi} | \uparrow_z \rangle
  |^2}}= {{ \cos^4(\theta/2)}\over{ \cos^4(\theta/2) +  \sin^2(\theta/2)}} 
\end{equation}
We have obtain two different results. This shows that this
interpretation of the ABL rule is incorrect.

\vskip .2cm
\noindent
{\bf Interpretation (c)~~}
{\it Counterfactual probability as the probability for the results of
  a measurement if it has been performed in the world ``closest'' to
  the actual world.}

This is identical in form and spirit to the theory of counterfactuals
of Bennett (1984), although the context of the pre- and post-selected
quantum measurements is somewhat beyond what he considered. This
interpretation is explicitly time-symmetric. The title, however, does
not specify it completely and I shall explain what do I mean (in
particular by the word ``closest'') now.

I have to specify the concept of ``world''. There are many parts of
the world which do not interact with the quantum system in question,
so their states are irrelevant to the result of the measurement. In
our discussion we might include all these irrelevant parts, or might
not, without changing any of the conclusions.  There are other aspects
of the world which are certainly relevant to the measurement at time
$t$, but we postulate that they should be disregarded. Everything
which is connected to our decision to perform the measurement at time
$t$ and all the records of the result of that measurement are not
considered.  Clearly, the counterfactual world in which a certain
measurement has been performed is different from an actual world in
which, let us assume, no measurement has been performed at time
$t$. The profound differences are both in the future where certain
records exist or do not exist and in the past which must be different
since one history leads to performing the measurement at time $t$ and
another history leads to no measurement.\footnote{If a random process
chooses between the two possibilities, then the past before this
process might be identical.}  However, our decision to make the
measurement is not connected to the quantum theory which makes
predictions about the result of that measurement. We want to limit
ourselves to the discussion of the time-symmetry of the quantum
theory. We do not consider here the question of the time-symmetry of
the entire world. Therefore, we exclude the external parts from our
consideration.

What constitutes a description of a quantum system itself is also a
very controversial subject.  The reality of the Schr\"odinger wave,
the existence or inexistence of hidden variables etc. are subjects of
hot discussions. However, everybody agrees that the collection of all
results of measurements is a consistent (although maybe not complete)
description of the quantum system. Thus, I propose the following
definition: \begin{quotation} {\em A world ``closest'' to the actual
world is the world in which all measurements (except the measurement
at the time $t$ if performed) have the same outcomes as in the actual
world.}  \end{quotation} This definition overcomes the common
objection according to which one should not consider together
statements about pre- and post-selected systems regarding different
measurements at time $t$ because these systems belong to different
ensembles. The difference is in their quantum state at the time period
between $t$ and $t_2$.\footnote{If one is adopting our backward
evolving quantum state, he can add that the systems are also different
due to the backward evolving state between $t$ and $t_1$.} Formally,
the problem is solved by considering only results of measurements and
not the quantum state.  The justification of this step follows from
the rules of the game: it is postulated that the quantum system is not
disturbed during the periods of time $(t_1, t)$ and $(t,
t_2)$. Therefore, it is postulated that no measurement on the system
is performed during these periods of time. Since unperformed
measurements have no results, the difference between the ensembles has
no physical meaning in the discussed problem.

From the alternatives I presented here, only interpretation (c) is
time-symmetric. This is the reason why I believe that it is the only
reasonable candidate for analyzing the (time-symmetric) problem of
measurements performed between two other measurements.

A very serious study of time's arrow and counterfactuals, in
particular, in the framework of quantum theory, was performed recently
by Price (1996).  Let me quote from his section ``Counterfactuals:
What should we fix?'': \begin{quotation} Hold fixed the past, and the
same difficulties arise all over again. Hold fixed merely what is
accessible, on the other hand, and it will be difficult to see why
this course was not chosen from the beginning. (1996, 179)
\end{quotation} This quotation looks very much like my proposal. And
indeed, I find many arguments in his book pointing in the same
direction. However, this quotation represents a time asymmetry:
``merely what is accessible'' is, in fact, ``an accessible past''. But
this is not the time asymmetry of the physical theory; Price writes:
``no physical asymmetry is required to explain it.''  Although the
books includes an extensive analysis of a photon passing through two
polarizers -- the classic setup for the ABL case, I found no explicit
discussion of a possible measurement in between, the problem we
discuss here.\footnote{Price briefly and critically mentions the ABL
paper. He
  writes (1996, 208): ``What they [ABL] fail   to note, however, is
  that  their argument  does nothing to address the problem for those
 who disagree with Einstein -- those  who think that the state
 function is a complete description, so that the change that takes
 place on measurements is a real change in the world, rather than
 merely change in our knowledge of the world.'' This seems to me an
 unfair criticism: the ABL clearly state that in the situations they
 consider ``the complete description'' is given by  {\it two} wave
 functions, see more in Aharonov and Vaidman (1991). Moreover, it seems
 to me that the  development of this time-symmetric quantum formalism
 is not too far from the spirit of the ``advanced action'' -- the Price
 vision of the solution of the time's arrow problem.}

\vskip 1.cm \noindent
{\bf 5. Inconsistency proofs.~~}
The key point of the criticism of the time-symmetrized quantum theory
is the conflict between counterfactual interpretation of the ABL rule
and predictions of quantum theory. I shall argue here that the proofs
of the inconsistency are unfounded and therefore the criticism
essentially falls apart.

The inconsistency proofs (Sharp and Shanks 1993; Cohen 1995; Miller
1996) have the same structure.  Three consecutive measurements are
considered. The first is the preparation of the state $|\Psi_1\rangle$
at time $t_1$.  The probabilities for the results $a_i$ of the second
measurement at time $t$ are considered. And the final measurement at
time $t_2$ is introduced in order to allow the analysis which uses the
ABL formula.  Sharp and Shanks consider three consecutive spin
component measurements of a spin-1/2 particle in different
directions. Cohen analyses a particular single-particle interference
experiment. It is a variation on the theme of Mach-Zehnder
interferometer with two detectors for the final measurement and the
possibility of placement of a third detector for the intermediate
measurement. Finally, Miller repeated the argument for a
 system of tandem Mach-Zehnder interferometers.
  In all cases the ``pre-selection only''
situation is considered.

It is unnatural to apply the time symmetrized formalism for such
cases. However, it must be possible.  Thus, I should not show that the
time-symmetrized formalism has an advantage over the standard
formalism for describing these situations; I should only show
consistency. In the standard approach to quantum theory the
probability for the result of the measurement of $A$ at  time $t$
is given by Eq. 1. The claim of all the proofs is that the counterfactual
interpretation of the ABL rule yields a different result.  In all cases
the final measurement at time $t_2$ has two possible outcomes
which we signify as ``$1_f$'' and ``$2_f$''; they are spin ``up'' or
``down'' and the click of the detector $D_1$ or $D_2$ respectively.
The suggested application of the ABL rule is as follows. The
probability for the result $a_i$ is:
\begin{equation}
  \label{p1}
  {\rm Prob}(A=a_i)=  {\rm Prob}(1_f)  {\rm Prob}(A=a_i |1_f) +
  {\rm Prob}(2_f)  {\rm Prob}(A=a_i |2_f),
\end{equation}
where $ {\rm Prob}(A=a_i |1_f)$ and $ {\rm Prob}(A=a_i |2_f)$ are the 
conditional probabilities given by the ABL formula, Eq. 2, and $ {\rm
  Prob}(1_f)$ and $ {\rm Prob}(2_f)$ are the probabilities for the
results of the final measurement.  There is no ambiguity about the
probability of the intermediate measurement given the result of the
final measurement, it is uniquely defined by the ABL formula. The
error in the proofs is in the calculation of the probabilities $ {\rm
  Prob}(1_f)$ and $ {\rm Prob}(2_f)$ of the final measurement. In all three
cases it was calculated on the assumption that {\rm no} measurement
took place at  time $t$. Clearly, one cannot make this assumption
here since then the discussion about the probability of the result of
the measurement at  time $t$ is meaningless. Unperformed
measurements have no results. Thus, there is no  surprise that the
value for the probability ${\rm Prob}(A=a_i)$ obtained in this way
comes out different from the value predicted by the quantum theory.

Straightforward calculations show that if one uses the formula (6)
with the probabilities $ {\rm Prob}(1_f)$ and $ {\rm Prob}(2_f)$
calculated on the condition that the intermediate measurement has been
performed, then the outcome is the same as predicted by the standard
formalism of quantum theory.  Consider, for example, the experiment
suggested by Sharp and Shanks, the consecutive spin measurements with
the three directions in the same plane and the relative angles
$\theta_{ab}$ and $\theta_{bc}$. The probability for the final result
``up'' is

\begin{equation}
  \label{p1f}
  {\rm Prob}(1_f) =
\cos^2(\theta_{ab}/2)\cos^2(\theta_{bc}/2)+
\sin^2(\theta_{ab}/2)\sin^2(\theta_{bc}/2),
\end{equation}
  and the probability for
the final result ``down'' is
\begin{equation}
  \label{p12}
 {\rm Prob}(2_f)
=\cos^2(\theta_{ab}/2)\sin^2(\theta_{bc}/2)+
\sin^2(\theta_{ab}/2)\cos^2(\theta_{bc}/2). 
\end{equation}
 The ABL formula yields
\begin{equation}
  \label{p1abl}
 {\rm Prob}(up |1_f)={{\cos^2(\theta_{ab}/2)\cos^2(\theta_{bc}/2)}
  \over {\cos^2(\theta_{ab}/2)\cos^2(\theta_{bc}/2)+
    \sin^2(\theta_{ab}/2)\sin^2(\theta_{bc}/2)}}
\end{equation}
 and 
\begin{equation}
  \label{p2abl}
 {\rm
  Prob}(up |2_f)={{\cos^2(\theta_{ab}/2)\sin^2(\theta_{bc}/2)}
  \over {\cos^2(\theta_{ab}/2)\sin^2(\theta_{bc}/2)+
    \sin^2(\theta_{ab}/2)\cos^2(\theta_{bc}/2)}}.
\end{equation}
 Substituting all
these equations into Eq. 6 we obtain
\begin{equation}
  \label{paiabl}
 {\rm
  Prob}(up)=\cos^2(\theta_{ab}/2).
\end{equation}
This result coincide with the prediction of the standard quantum
theory. It is a straightforward exercise to show in the same way that
no inconsistency arises also in the examples of Cohen and Miller.
    
Apparently, the motivation of the authors of the above inconsistency
proofs for taking the expressions for the probabilities $ {\rm
  Prob}(1_f)$ and $ {\rm Prob}(2_f)$ based on the assumption that no
measurement has been performed at time $t$ follows from their
interpretation of ``counterfactual interpretation of the ABL rule''
which was named (a) above. It seems that they consider that a
necessary condition for a counterfactual is that it {\em has} to be
contrary to what is in the actual world. In their view the only
alternative to the postulate of ``no measurement'' is the postulate
that a measurement has been actually performed.\footnote{In this case
  the authors of the inconsistency proofs say that no contradiction
  arises, but also no interesting question can be asked.} I believe,
however, that one can interpret counterfactuals without postulating
that they are necessarily contrary to the actual world, see
interpretation (c) above.  Moreover, as I explained above, I find
their interpretation (a) physically meaningless.

\vskip 1.cm \noindent
{\bf 6. Time asymmetry prejudice.~~}
In my approach the pre- and post-selected states are given. Only
intermediate measurements are to be discussed. So the frequently posed
question about the probability of the result of the post-selection
measurement is irrelevant.  It seems to me that the critics of the
time-symmetrized quantum theory use in their arguments the
preconception of an asymmetry. It is not surprising then that they
reach various contradictions.  Probably the first to go according to
this line were Bub and Brown:
\begin{quotation}
Put simply, systems initially in the state $\psi_I$ which are subject
to an $N$ measurement, and subsequently yield the state $\psi_F$ after
an $M_F$ measurement, would not necessarily yield this final state if
subjected to a measurement of $M$ instead of $N$. (1986, 2338)
\end{quotation}
Their argument is valid (see Albert et al. 1986) in
the context of the possible hidden variable theories which allow us to {\em
  predict} the results of measurements, but it should not be brought
against proposals of a time-symmetrized formalism as it was done
frequently later.  Let me quote a few examples: Cohen  writes:
\begin{quotation}
We have no reason to expect that, for example, the $N/4$ systems
preselected by $|\psi_1(t_1)\rangle$ and post-selected by
$|\psi_2(t_2)\rangle$ after an intermediate measurement of
$\sigma_{1y}$ would still have yielded the state  $|\psi_2\rangle$
after an intermediate measurement of $\sigma_{2x}$ or of
$\sigma_{1y}\sigma_{2x}$ instead of $\sigma_{1y}$.(1995, 4375)    
\end{quotation}
A consistent time-symmetric approach should question the pre-selection
on the same footing as the post-selection; or rather not question any of
them, as I propose.

Another asymmetry pre-conception lead to the ``retrodiction paradox''
of Peres:
\begin{quotation}
The asymmetry between prediction and retrodiction is related to the
fact that predictions can be verified (or falsified) by actual
experiments, retrodictions cannot. Retrodictions are counterfactual
statements about {\em unperformed experiments}; in quantum mechanics,
unperformed experiments have no results (Peres 1987). (1994, 23)
\end{quotation}
While I certainly agree that unperformed experiments have no results,
I challenge the interpretation according to which counterfactual
statements are  necessarily about events which do not happen. The
asymmetry considered by Peres (1994) is, in fact, between {\em
  prediction} based on the results in the past of time $t$ in
question and {\em inference} based on the results both in the past and
in the future of  time $t$. This inference he erroneously considered
as retrodiction (see Aharonov and Vaidman 1995).

If we are not considering a pre- and post-selected system then there
is an asymmetry between prediction and retrodiction.  For example
(Aharonov and Vaidman 1990, 11-12), assume that the $ x$ component of
the spin of a spin-1/2 particle was measured at time $t$, and was
found to be $\sigma_x =1$.  While there is a symmetry regarding
prediction and retrodiction for the result of measuring $\sigma_x$
after or before time $t$ (in both cases we are certain that $\sigma_x
=1$), there is an asymmetry regarding the results of measuring
$\sigma_y$. We can predict equal probabilities for each outcome,
$\sigma_y  = \pm 1$, of a measurement performed after time $t$, but we
cannot claim the same for the result of a measurement of $\sigma_y$
performed before time $t$.  The difference arises from 
 the usual assumption that there is no
``boundary condition'' in the future, but there is a boundary
condition in the past: the state in which the particle was prepared
before time $t$. Maybe in a somewhat artificial way we can reconstruct
the symmetry even here, out of the context of pre- and post-selected
systems. We can ``erase'' the results of the measurements of the spin
measurements in the past (Vaidman 1987, 61). In order to do this we
perform at time $t_0$, before time $t$, a measurement of a Bell-type
operator on our particle and another auxiliary particle, an  ancilla.
We ensure that no measurement is performed on the particle between
$t_0$ and $t$ (except the possible measurement whose result we want to
consider) and we prevent any measurement on the ancilla  from
time $t_0$ and on. 
 The Bell-type measurement
correlates the quantum state of our particle evolving from the past
with the results of the future measurement performed on the ancilla.
Since the latter is unknown, we obtain, effectively, an unknown past for
our particle. Now, for such a system, if we know that the result of the
measurement at  time is $\sigma_x =1$, we can also {\em
  retrodict} that there are equal probabilities for both outcomes of
the measurement of $\sigma_y$ performed before time $t$ (but after
time $t_0$).  The time symmetry is restored.\footnote{The time symmetry
  is restored not just for the $\sigma_y$ measurement, but for any
  spin measurement.}

\vskip 1.cm \noindent {\bf 7. Time symmetry of the process of
measurement.~~} Obviously, in order to discuss a measurement at time
$t$ between two other measurements in a time-symmetric way, the
process of measurement at time $t$
 must  be time-symmetric. Usually, a 
measurement of a quantum variable $A$  is modeled by the von Neumann
(1955) Hamiltonian 
\begin{equation}
  \label{neumann}
 H = g(t) p A,
\end{equation}
where $p$ is the momentum conjugate to the pointer variable $q$, and
the normalized coupling function $g(t)$ specifies the time of the
measurement interaction. The function $g(t)$ can be made symmetric in
time (not that it matters) and the form of the coupling then is
time-symmetric. The result of the measurement is the difference
between the value of $q$ before and after the measurement interaction.
So it seems that everything is time-symmetric.

However, usually there is an asymmetry in that that the initial position
of the pointer is customized to be zero (and therefore the final
position correspond to the measured value of $A$). This seemingly
minor aspect points  to a genuine asymmetry. Of course, the initial
zero position of the pointer  is  not
a necessary condition; we can choose any other initial position as well.
But, we cannot chose the
final position.   We {\em know} the initial position and we
find out, at the end of the measurement the final position. We can
introduce another step with symmetrical coupling, but we will not be
able to remove the basic asymmetry: we do not know the result of the
measurement before the measurement but we do know it after the
measurement.

This asymmetry in time is an intrinsic property of the concept of
measurement and it has no connection to the quantum theory.  It is
related to the arrow of time based on the increasing memory. See
illuminating discussion of Bitbol (1988) of the process of measurement
in the framework of the many-worlds interpretation  (Everett, 57).

The symmetry aspect of the process of measurement which is important
for our discussion is that a measurement at time $t$ leads to
identical forward and backward evolving states out of time $t$.
Operationally, it means that under the assumption of zero Hamiltonian
and and identical pre- and post-selected states (or mixtures), the
probabilities for the result of any measurement performed before $t$
(but after the pre-selection) is equal to that performed after time
$t$ (but before the post-selection).  The measurement described by the
Hamiltonian (12) has this time symmetry.  Moreover, any ``ideal'' von
Neumann measurement, which projects on the property to be measured and
does not change the quantum state if it has the measured property, is
symmetric in this sense. Recently Shimony (1995) proposed considering
more general quantum measurements which do not change the measured
property (so they are repeatable, the main property which is required
from a ``good'' measurement) but which change the state (even if it
has the measured property). Such measurements are intrinsically
asymmetric in our sense, since given identical boundary conditions in
the past and in the future they lead to different probabilities for
the results of some measurements performed before or after the time of
the generalized measurement. Clearly, the time symmetrized formalism
is not applicable for such measurements as Shimony has showed.

\vskip 1.cm \noindent
{\bf 8. Elements of Reality.~~}
Until now we have discussed a situation in which we know the results
of the measurements at   $t_1$ and $t_2$, we know that there is
no measurement or disturbance of the system between $t_1$ and $t_2$
except, may be, a measurement at  time $t$. We  also discussed the
outcomes of that possible measurement. For the same system we can
discuss several, in general incompatible, possible measurements at
time $t$ and this is why we consider it as  counterfactual
reasoning. Interesting novel (relative to a pre-selection only
situation) structures emerge in this situation. In particular, there
are situations in which several incompatible measurement (if
performed) have certain results with probability 1. I have proposed to
call them {\em elements of reality} (Vaidman 1993a, 1996). These
elements of reality, contrary to the pre-selected situations might
contradict the {\em product rule} (Vaidman 1993b), i.e., if $A=a$ and
$B=b$ are elements of reality, $AB =ab$ might not be an element of
reality.

One important aspect of these elements of reality is that they are
Lorentz invariant. Their introduction solves an 
apparent contradiction which follows from the
existence of Lorentz invariant elements of reality.  Let me quote a
recent paper by Cohen and Hiley:
\begin{quotation}
  A further criticism by Vaidman (1993) leads to the conclusion that
  the gedanken experiment [of Hardy, 1992] does not lead to any
  contradiction, because it involves a pre- and post-selected quantum
  system, for which, it is claimed, the ``product rule'' does not
  apply. Unfortunately, as we show in a separate paper [apparently
  Cohen and Hiley 1996], Vaidman's analysis is not valid because it
  makes incorrect use of the formula of Aharonov, Bergman and Lebowitz
  (1964; Aharonov and Vaidman 1991).(1995, 76)
\end{quotation}
I believe that I have succeeded here in showing the legitimacy of
applying the ABL formula for pre- and post-selected quantum systems
with a choice of possible intermediate measurements which is exactly
the situation considered in Hardy's gedanken experiment and thus
defending my results (1993).

One might argue about the significance of these concepts beyond the
philosophical construction of (jointly unmeasurable) ``elements of
reality'', since it is impossible to perform incompatible measurements
on a single system. I find the most important aspect of these concepts
in their relation to {\em weak measurements} (Aharonov and Vaidman
1990).\footnote{The scope of this paper allows only to touch this
  broad issue, but one can find more on this subject in Vaidman
  (1996) where, in particular, I defined a new type of reality based on
  the concept of weak measurements.} Weak measurements are almost standard
measurement procedures with weakened coupling. Weak measurements
essentially do not change the quantum states (evolving forward and
backwards in time) of the system.  Several weak measurements can be
performed on a single system and they are compatible even though their
counterparts, the ideal measurements are not compatible.

An example of an interesting connections between weak and strong
(ideal) measurements is the theorem (Aharonov and Vaidman 1991) which
says that if the probability for a certain value to be the result of a
strong measurement is 1, then the corresponding weak measurement must
yield the same value.\footnote{In such situation, when we discuss the
  outcomes of weak measurements, the statements about strong
  measurement have even a stronger counterfactual sense.  The
  counterfactual reasoning about {\it unperformed} strong measurements
  (which correspond to elements of reality) helps us to find out the
  results of the {\em performed} weak measurements.} However, in
general, the outcome of weak measurements might not be one of the
possible outcomes of a strong measurement. The outcome of the weak
measurement of a variable $A$ is the weak value (Eq. 3) which might
lie far away from the range of the eigenvalues of $A$. The weak value
is not just a theoretical concept related to a gedanken experiment.
Recently, weak values have been measured in a real laboratory (Ritchie
et al., 1991).

\vskip 1.cm \noindent {\bf 9. Conclusions.~~} In this paper I have
defended the time-symmetrized quantum formalism originated by Aharonov
et al. (1964) against recent criticism.  The criticism followed from
the pre-conception of time asymmetry which is the feature of the
standard formalism of quantum theory and of the standard approach to
counterfactual reasoning. I have argued, that in the context of the
experiments on pre- and post-selected quantum systems, the
time-symmetric counterfactual theory suggested by Bennett (1984) is
the most appropriate. I introduced the time-symmetric counterfactual
interpretation of the ABL rule and showed that it does not lead to any
contradiction with the predictions of quantum theory.

I disagree in an essential way with a large number of recent works. In
particular, contrary to the conclusions of Sharp and Shanks (1993), I
believe that the time-symmetrized quantum theory ``yields fresh
insights about the fundamental interpretive issues in quantum
mechanics''.  I base my belief on the research in which I took part
and which allowed us to see numerous surprising quantum effects which are
hidden in the framework of the standard approach in  very complicated
mathematics of some peculiar interference effects (e.g. Aharonov et
al. 1987, 1990, 1993; Vaidman 1991). I see a novel rich structure in
time-symmetrized quantum theory which suggests and solves
surprising quantum problems  (Vaidman et al. 1987; Vaidman 1996).
It is also plausible that the time-symmetrized approach might help
investigated the current problems of quantum gravity (see, for
example, Unruh 1995).

The time-symmetrized quantum theory fits well into the many-worlds
interpretation (MWI), my preferred interpretation of quantum theory
(Vaidman 1993c, 1994). The counterfactual worlds corresponding to
different outcomes of quantum measurements have in the MWI an
especially clear meaning: these are subjectively actual different
worlds. In each world the observers of the quantum measurement call
their world as actual, but, if they believe in the MWI they have no
paradoxes about ontology of the other worlds.  Consider an
illuminating example (this time of pre-selected only situation) by
Mermin (1989) in which counterfactual reasoning lead him to wonder:
can he help his favorite baseball team to win by watching their game
on television?  Or, can an action in one region change something in a
space-time separated region? The counterfactual reasoning in the
framework of the standard (single-world) interpretations lead him to
the paradoxical answer ``yes''.  The MWI answers that his action in
one place causes different separation into worlds which include
correlations between the two regions and therefore we have, this time
a not surprising answer ``yes''. If we consider the problem from an
external position, i.e. we consider the whole physical universe which
incorporates all the worlds, then we obtain the expected answer
``no''.\footnote{The answer ``no'' is expected because of the locality
of physical interactions.}  Some things at the remote location become
correlated to different things at the first location, but it does not
change any measurable property in the remote location.

The MWI interpretation yields also a convincing answer to paradoxical
situations considered by Penrose: \begin{quotation}
  \noindent What is particularly curious about quantum theory is that
there can be actual physical effects arising from what philosophers
refer to as {\em counterfactuals} -- that is, things that might have
happened, although they did not happened. (1994, 240) \end{quotation}
According to the MWI, in the situations considered by Penrose,
``things'' did not happened in a particular world, but did happened in
some other world (see Vaidman 1994). Therefore, they did took place in
the physical universe and thus their effect on some other facts in the
physical universe is not so surprising.

It is a pleasure to thank Yakir Aharonov, David Albert, Avshalom
Elitzur, Lior Goldenberg, Yoav Ben-Dov, Igal Kvart, Abner Shimony and
Steve Wiesner for helpful discussions.  The research was supported in
part by grant 614/95 of the Basic Research Foundation (administered by
the Israel Academy of Sciences and Humanities).

\vskip .8cm
\vfill
\break

 \centerline{\bf REFERENCES}
\vskip .15cm
\footnotesize

\vskip .13cm \noindent 
Aharonov, Y. and Albert, D. (1987), 
``The Issue of Retrodiction in Bohm's Theory'',
in B.J. Hiley and F.D. Peat (eds.) {\em Quantum Implications}. New
York: Routledge \& Kegan Paul, pp.224-226.

\vskip .13cm \noindent 
 Aharonov, Y., Albert, D., Casher, A., and Vaidman, L. (1987),
 ``Surprising Quantum Effects'',
{\em  Physics Letters A 124}: 199-203.

\vskip .13cm \noindent 
Aharonov, Y.,  Albert, D., and D'Amato, S. (1985),
``Multiple-Time Properties of Quantum Mechanical Systems'',
 {\em Physical Review  D 32}:  1975-1984.

\vskip .13cm \noindent 
Aharonov, Y., Anandan, J., Popescu, S., and Vaidman, L.  (1990),
``Superpositions of Time Evolutions of a Quantum System and a Quantum Time
Machine'',
{\em  Physical Review Letters 64}: 2965-2968.
 
\vskip .13cm \noindent 
Aharonov, Y.,  Bergmann,  P.G., and  Lebowitz, J.L. (1964),
``Time Symmetry in the Quantum Process of Measurement'',
 {\em Physical Review 134B}: 1410-1416. 

\vskip .13cm \noindent 
Aharonov, Y.,  Popescu, S., Rohrlich, D., and Vaidman, L.  (1993),
 ``Measurements, Errors, and Negative Kinetic Energy''
 {\em Physical   Review A 48}: 4084-4090.
 
\vskip .13cm \noindent 
Aharonov, Y. and Vaidman, L. (1990),
 ``Properties of a Quantum System
During the Time Interval Between Two Measurements'',
{\em Physical Review A 41}: 11-20.

\vskip .13cm \noindent 
Aharonov, Y. and Vaidman, L. (1991),
``Complete Description of a Quantum System at a Given Time'',
{\em Journal of  Physics  A 24}: 2315-2328.

\vskip .13cm \noindent Aharonov, Y. and Vaidman, L. (1995), ``Comment
on `Time Asymmetry in Quantum Mechanics: a Retrodiction Paradox' '',
{\em Physical Letters A 203}:148-149.

\vskip .13cm \noindent 
Albert, D. (1992), {\em Quantum Mechanics and Experience},
 Cambridge:Harvard University Press.

\vskip .13cm \noindent 
Albert, D.,  Aharonov, Y. and  D'Amato, S. (1986),
``Comment on `Curious Properties of Quantum Ensembles which have been Both
  Preselected and Post-Selected' '',
{\em Physical Review Letters 56}:2427.

\vskip .13cm \noindent 
Bedford, D. and Stapp, H.P. (1995),
``Bell's Theorem in an Indeterministic Universe'',
{\em Synthese 102}: 139-164.

\vskip .13cm \noindent 
Bennett, J. (1984),
``Counterfactuals and Temporal Direction''
{\em Philosophical Review 93}: 57-91.

\vskip .13cm \noindent 
Bitbol, M. (1988),
``The Concept of Measurement and  Time-Symmetry in  Quantum Mechanics'',
{\em Philosophy of Science 55}:349-375.

\vskip .13cm \noindent 
 Bohm, D. (1952),
 ``A Suggested Interpretation of the Quantum Theory in Terms of
`Hidden' Variables I and II'', {\it Physical Review  85}: 97-117.

\vskip .13cm \noindent 
Bohm, D. and Hiley, B. J. (1993),
{\em The Undivided Universe: An Ontological Interpretation of Quantum
  Mechanics},
London:Routledge.

\vskip .13cm \noindent 
 Bub, J. and Brown, H. (1986), 
``Curious Properties of Quantum Ensembles which have been Both
  Preselected and Post--Selected'',
{\em Physical Review Letters 56}: 2337-2340.

\vskip .13cm \noindent 
Cohen,  O. (1995), 
``Pre- and Post-selected Quantum Systems, Counterfactual Measurements,
and Consistent Histories'',
 {\em Physical Review  A 51}: 4373-4380. 
 
\vskip .13cm \noindent 
 Cohen, O. and Hiley, B.J. (1995),
  ``Reexamining the Assumptions that Elements of Reality can be
  Lorentz Invariant'', {\em Physical Review A 52}: 76-81.

\vskip .13cm \noindent 
 Cohen, O. and Hiley, B.J. (1996), 
`` Elements of Reality, Lorentz Invariance and the Product Rule'',
 {\em Foundations of Physics 26}: 1-15.

\vskip .13cm \noindent 
Everett, H. (1957),
 `` `Relative State' Formulation of Quantum Mechanics'',
{\em Review of Modern Physics 29}: 454-462.

\vskip .13cm \noindent 
Griffith, R. B. (1984),
``Consistent Histories and the Interpretation of Quantum Mechanics'',
{\em Journal of Statistical Physics 36}: 219-272.

\vskip .13cm \noindent Hardy, L. (1992) ``Quantum Mechanics, Local
Realistic Theories, and Lorentz-Invariant Realistic Theories'', {\em
Physical Review Letters 68}:2981-2984.

\vskip .13cm \noindent 
Lewis, D. (1973),
{\em Counterfactuals},
Oxford:Blackwell Press.

\vskip .13cm \noindent 
 Lewis, D. (1986)
 ``Counterfactual Dependence and
  Time's Arrow'' reprinted from {\em Nous 13}: 455-476 (1979) and {\em
    Postscripts to ``Counterfactual ..''} in Lewis, D. {\em
    Philosophical Papers Vol.II}, Oxford: Oxford University Press,
  pp.32-64.

\vskip .13cm \noindent 
Mermin, D. (1989),
``Can You Help Your Team Tonight by Watching on TV? More Experimental
Metaphysics from Einstein, Podolsky, and Rosen'',
in J.T. Cushing and E. McMullin (eds.) {\em Philosophical Consequences
  of Quantum Theory: Reflections on Bell's Theorem.} Notre Dame:
University of Notre Dame Press, pp.38-59.

\vskip .13cm \noindent 
Miller, D. J. (1996),
``Realism and Time Symmetry in Quantum Mechanics'',
{\em Physical Letters A}, to be published.

\vskip .13cm \noindent 
Penrose, R. (1994),
{\em Shadows of the Mind},
 Oxford: Oxford University Press.

\vskip .13cm \noindent 
Peres, A. (1978),
``Unperformed Experiments Have no Results'',
{\em American Journal of Physical 46}: 745-747.

\vskip .13cm \noindent 
Peres, A. (1993),
{\em Quantum Theory:Concepts and Methods},
Dordrecht: Kluwer Academic Publisher.

\vskip .13cm \noindent 
Peres, A. (1994),
``Time Asymmetry in Quantum Mechanics: a Retrodiction Paradox'',
{\em Physical Letters A 194}:21-25.
 
\vskip .13cm \noindent 
Price H. (1996),
{Time's Arrow}, New York: Oxford University Press.

\vskip .13cm \noindent 
Ritchie, N.W.M.,  Story, J.G. and  Hulet, R.G. (1991),
``Realization of a `Measurement of a Weak Value''',
{\em Physical Review Letters 66}: 1107-1110.

\vskip .13cm \noindent 
Sharp, W.D. and Shanks, N. (1993),
``The Rise and Fall of Time-Symmetrized Quantum Mechanics'',
{\em Philosophy of Science 60}:488-499.

\vskip .13cm \noindent 
 Shimony, A. {1995}
``A Bayesian Examination of Time-Symmetry in the Process of
Measurement'' Preprint of Boston University.

\vskip .13cm \noindent 
 Skyrms, B. (1982),
 ``Counterfactual Definetness and Local
 Causation'', {\em Philosophy of Science 49}:43-50.

\vskip .13cm \noindent 
Vaidman, L. (1987),
``The problem of the interpretation of relativistic quantum
theories'',
{\em Ph.D. Thesis}, Tel-Aviv University.

\vskip .13cm \noindent 
 Vaidman, L., Aharonov, Y., Albert, D., (1987), ``How to Ascertain the
Values of $\sigma_x, \sigma_y,$ and $\sigma_z$ of a Spin-1/2
Particle'' {\em Physical Review Letters 58}: 1385-1387.

\vskip .13cm \noindent 
 Vaidman, L. (1991),
``A Quantum Time Machine''
{\em Foundation of  Physics  21}: 947-958.

\vskip .13cm \noindent 
 Vaidman, L. (1993a),
``Lorentz-Invariant `Elements of Reality' and the Joint
Measurability of Commuting Observables'',
{\em Physical Review Letters 70}: 3369-3372.

\vskip .13cm \noindent 
Vaidman, L. (1993b),
`` `Elements of Reality' and the Failure of the Product Rule''
  in P.J. Lahti, P.~Bush, and P. Mittelstaedt (eds.),{\em Symposium on
the Foundations of Modern Physics}. New Jersey: World Scientific,
pp. 406-417.

\vskip .13cm \noindent 
Vaidman, L. (1993c),
``About Schizophrenic Experiences of the Neutron or Why We
should Believe in the Many-Worlds Interpretation of Quantum Theory,'' 
Tel-Aviv University Preprint TAUP 2058-93.

\vskip .13cm \noindent 
 Vaidman, L. (1994),
 ``On the Paradoxical Aspects of New Quantum Experiments''
 {\it Philosophy of Science Association 1994} pp. 211-217.

\vskip .13cm \noindent 
Vaidman, L. (1996),
``Weak-Measurement Elements of Reality'',
 {\em Foundations of Physics 26?}:895-906.

\vskip .13cm \noindent 
von Neumann, J. (1955),
  {\em Mathematical Foundations of Quantum Theory},
  Princeton: Princeton University Press.

\vskip .13cm \noindent 
Unruh, W. (1995),
``Time, Gravity, and Quantum Mechanics.''
in S. F. Savitt (ed.), {\em Time's Arrow Today}.
 Cambridge: Cambridge University Press.

\end{document}